\documentclass[aip,rsi,amsmath,amssymb,reprint,longbibliography]{revtex4-2}

\usepackage{graphicx}
\usepackage{dcolumn}
\usepackage[separate-uncertainty = true,multi-part-units=single]{siunitx}
\usepackage{lipsum}

\usepackage[utf8]{inputenc}
\usepackage[T1]{fontenc}

\usepackage{textcomp}
\usepackage{hyperref}

\makeatletter
\def\@email#1#2{%
 \endgroup
 \patchcmd{\titleblock@produce}
  {\frontmatter@RRAPformat}
  {\frontmatter@RRAPformat{\produce@RRAP{*#1\href{mailto:#2}{#2}}}\frontmatter@RRAPformat}
  {}{}
}%
\makeatother
\begin{document}

\preprint{AIP/123-QED}

\title[CHIMERA]{CHIMERA: Couette Hydrodynamics for the Investigation of Multi-fluid Extreme-Reynolds Turbulence in Annuli}

\author{Pantxo Diribarne}
\email{pantxo.diribarne@univ-grenoble-alpes.fr}
\author{Jérôme Chartier}
\author{Jérôme Duplat}
\author{Bernard Rousset}
\affiliation{dSBT/IRIG CEA, Universit\'{e} Grenoble Alpes - F-38054
  Grenoble, France}

\date{\today}
 \DeclareSIUnit\bar{bar}
\begin{abstract}
The CHIMERA (Couette Hydrodynamics for the Investigation of
Multi-fluid Extreme-Reynolds Turbulence in Annuli) apparatus has been
developed to investigate turbulent momentum transport in a canonical
wall-driven flow using a wide variety of working fluids, from
room-temperature gases to normal and superfluid helium. We present the
design and characterization of a cryogenic Taylor-Couette facility in
which torque is measured on a suspended outer cylinder using a
torsion-pendulum sensor and a non-contact optical readout. This
configuration avoids the need for cryogenic torque transducers or
electrical instrumentation inside the cryostat while providing high
sensitivity over a wide dynamic range. Measurements performed in
gaseous helium at room temperature and in normal liquid helium reach
Reynolds numbers up to $10^6$ and agree well with established torque
correlations, validating the apparatus over more than five decades in
Reynolds number. Beyond providing such an exceptionally broad
operating range with a single experimental facility and a single
torque sensor, the principal motivation of CHIMERA is to enable direct
comparisons of turbulent momentum transport in He I and He II under
identical boundary conditions. By combining the same geometry and the
same global torque diagnostic in both classical and quantum fluids,
the apparatus provides a unique platform for investigating how
wall-driven turbulence and its scaling evolve across the superfluid
transition.

\end{abstract}

\maketitle

\section{\label{sec:intro}Introduction}

Quantum turbulence has been investigated in a wide variety of flow
configurations \cite{VinenNimiela2002}. Thermal counterflow, which is
unique to superfluid helium, has provided fundamental insight into
vortex dynamics and mutual friction \cite{Vinen1957,
  barenghi2001quantized}. In contrast, experiments in inertially
forced flows, such as oscillating grids, propellers, or towed bodies,
have shown that many statistical properties of fully developed
turbulence remain remarkably close to those of classical fluids, with
the main differences appearing at the smallest scales
\cite{Maurer_1998, Stalp2002,  Salort_2012, Salort_2021,
  Bret25}. Comparatively little attention has been devoted to
wall-bounded shear flows, in which turbulence is generated exclusively
by the motion of smooth solid boundaries. Whether turbulent momentum
transport in such canonical flows remains governed by the same global
laws across the superfluid transition is still an open question. 

Taylor-Couette flow provides an attractive framework for addressing
this question. It is a canonical wall-bounded shear flow in which
turbulence is driven exclusively by the motion of smooth solid
boundaries, while the global transport of angular momentum can be
accurately quantified through torque measurements. Decades of
experimental and numerical investigations have made Taylor-Couette
flow one of the best-characterized turbulent shear flows \cite{
  Grossman16_tc_high_re, Hollerbach2023}, providing an ideal benchmark
against which measurements performed in superfluid helium can be
compared. Taylor-Couette apparatuses have been used since the early
days following the discovery of superfluid helium
\cite{Hollis53_taylor_couette_helium,heikkila1955viscosity,bendt1967attenuation,snyder1974rotating,donnelly1988flow},
mainly  at the laminar-turbulent transition, but never for studies at
high Reynolds numbers. In particular, there is still a lack of
research examining the ultimate regime of turbulence when the boundary
layer near a cylindrical surface becomes turbulent. One of the
objectives of this new apparatus is to compare the behavior of
superfluid helium and normal liquid helium (denoted respectively He~II and
He~I  here after) in this configuration. Our objective is therefore
not to develop a  Taylor-Couette facility solely to extend the accessible
Reynolds-number range, but rather to exploit this canonical
configuration to compare turbulent momentum transport in He~I and
He~II under identical forcing, identical boundary conditions, and
using the same global diagnostic.

The Taylor-Couette flow, i.e. the flow between two concentric
cylinders rotating independently about their common axis, is one of
the canonical configurations of fluid mechanics.  
Owing to its simple geometry -- well-defined boundary conditions and
rich sequence of hydrodynamic instabilities and turbulent states -- it
has long served as a model system for investigating fundamental
questions in hydrodynamic stability, transition, and turbulence.
Throughout this paper, we adopt the strain based Reynolds-number
convention\cite{Dubrulle05_stability_tc}
commonly used in modern Taylor-Couette
studies~\cite{Grossman16_tc_high_re}. When the outer cylinder is at
rest, like in our experiment, it is defined as

$$Re_s = \frac{2}{1+\eta} \frac{R_i\Omega_i\delta}{\nu},$$ where
$\eta = R_i/R_o$ is the ratio between the radii of the inner
and the outer cylinders, $R_i$ and $R_o$ respectively,
$\Omega_i$ is the angular rotation frequency of the inner cylinder,
$\delta = R_0-R_i$ is the gap between the cylinders, and $\nu$ is the
kinematic viscosity. Using this convention, the first centrifugal
instability occurs at a critical Reynolds number that depends on the
geometry and rotation rates of the cylinders. For the simplest
configuration, in which only the inner cylinder rotates while the
outer cylinder remains at rest, Taylor demonstrated that the laminar
Couette flow becomes centrifugally unstable above a critical rotation
rate, leading to the formation of axisymmetric toroidal vortices now
known as Taylor vortices~\cite{taylor1923stability}. 

Early experimental studies of concentric-cylinder flows actually
predate Taylor's celebrated stability analysis by several
decades. Motivated primarily by the determination of fluid viscosity,
pioneering experiments by Mallock~\cite{Mallock1889} and
Couette~\cite{Couette1890} already used torsion-pendulum arrangements
to measure the torque transmitted through the fluid. 
Although their primary objective was rheological characterization,
these studies also explored flow regimes where inertial effects became
significant. Taylor's seminal work~\cite{taylor1923stability}
subsequently established the theoretical framework for centrifugal
instability, transforming the concentric-cylinder apparatus into one
of the canonical systems for studying hydrodynamic stability and,
eventually, turbulence. Later Wendt \cite{wendt1933turbulente},
Donnelly \cite{Donnelly_Torque_1958, Donnelly1960_visual} and Coles
\cite{coles1965transition} documented the progressive sequence of
bifurcations leading from 
steady vortices to wavy vortices and eventually to time-dependent and
turbulent states. The comprehensive regime map obtained by
\textcite{andereck1986flow}  
provided a detailed classification of flow states
over wide parameter ranges and remains a reference for modern
Taylor-Couette facilities. These studies established Taylor-Couette
flow as a uniquely accessible laboratory realization of wall-bounded
shear turbulence with independently tunable control parameters. 

At sufficiently high Reynolds numbers, the Taylor-Couette flow becomes
fully turbulent while preserving a well-defined global control
parameter and a direct connection between the measured torque and the
radial transport of angular momentum. High-precision torque
measurements by Lathrop and collaborators revealed nontrivial scaling
behavior and transitions between distinct turbulent flow
regimes~\cite{Lathrop92_taylor_couette_long}. Building on these
experimental observations, Eckhardt \textit{et
  al.}~\cite{Eckhardt07_torque_scaling} established a close analogy
between turbulent Taylor-Couette flow and Rayleigh-Bénard convection,
highlighting the common role of global transport laws in these
canonical systems. More recently, experiments by 
\textcite{vangils2011torque}, in the Twente Turbulent Taylor-Couette
\cite{VanGils11_taylor_couette_twente} (T$^3$C hereafter),
have explored the so-called ultimate turbulence regime, in which
boundary layers become turbulent and the torque scaling approaches
asymptotic predictions. These results have reinforced the role of
torque as a global observable directly quantifying angular momentum
transport and dissipation in strongly nonlinear states.

Nevertheless, due to the energy consumption associated with very high Reynolds numbers, only a limited number of experiments have been conducted, those approaching $Re_s = 10^6$ and involving torque measurements are listed in Table \ref{TableManips}

\begin{table}
\caption{\label{TableManips} Comparison of existing Taylor-Couette facilities.}
\centering
\begin{ruledtabular}
\begin{tabular}{l||c|c|c}
Laboratory & Aspect ratio $\Gamma$ & Radius ratio $\eta$ & Max $Re_s$ \\
\hline\hline
Maryland \cite{paoletti2012angular}
& 11.5 & 0.7245 & \num{4.3e+06} \\

Twente T$^3$C \cite{VanGils11_taylor_couette_twente}
& 11.7 & 0.716 & \num{2.3E+06} \\
(config. 1) & & & \\

Twente T$^3$C \cite{VanGils11_taylor_couette_twente}
& 23.2 & 0.833 & \num{1.1E+06} \\
 (config. 2) & & & \\
Cottbus \cite{merbold2013torque}
& 20 & 0.5 & \num{1.0E+06} \\

Okinawa \cite{Butcher24_taylor_couette_okinawa}
& 14.9 & 0.747 & \num{8.0E+05} \\

CHIMERA
& 22 & 0.85 & \num{1.0E+06} \\
\end{tabular}
\end{ruledtabular}
\end{table}

We would also like to point out that the closer the radius ratio is to
one, the more difficult it is to achieve a high Reynolds number. With a
radius ratio $\eta\approx 0.85$ in our setup, we can compare our
results to those obtained in the T$^3$C using a configuration with a radius
ratio slightly less (0.833), for which a maximum Reynolds number of
\num{1.13e6} was achieved. 

Since these pioneering experiments, several distinct strategies have
been developed for torque measurements in Taylor-Couette
apparatuses. Early investigation by Wendt relied on a direct
mechanical balance: the outer cylinder,  
mechanically isolated, transmitted the hydrodynamic torque to a calibrated
lever arm, and the equilibrium moment was determined using known weights. 
This purely mechanical method provided a direct measurement of the 
fluid-induced torque, albeit with limitations arising from bearing friction
and alignment tolerances. In contrast, later experiments such as those of
\textcite{coles1965transition} inferred torque indirectly from the driving
power required to maintain constant angular velocity of the rotating inner
cylinder, after careful subtraction of mechanical losses. 
Modern high-precision facilities, including those used in the high-Reynolds
number studies of \textcite{Lathrop92_taylor_couette_long} , 
\textcite{VanGils11_taylor_couette_twente}, and \textcite{Butcher24_taylor_couette_okinawa}
typically employ in-line torque transducers or strain-gauge instrumentation
mounted on the rotating shaft. While these techniques achieve excellent
resolution under ambient conditions, they rely on torque transmission 
through rotating mechanical elements and such configurations can become
technically challenging when extended to extreme environments, 
particularly cryogenic temperatures. 
Thermal contraction, lubrication constraints, and mechanical coupling
through rotating shafts may introduce parasitic torques or compromise
measurement resolution. These limitations motivate alternative
strategies for torque detection that are compatible with a broader
range of working fluids.

In the present work, we describe a Taylor-Couette apparatus, called CHIMERA
(Couette Hydrodynamics for the Investigation of Multi-fluid Extreme-Reynolds turbulence in Annuli), in which
only the inner cylinder rotates, while the outer cylinder is suspended
as a precision pendulum and used as the torque-sensing element. The
fluid-induced torque is inferred from the angular deflection of the
outer cylinder about its suspension axis, providing a direct and
mechanically simple measurement principle. By eliminating torque
transmission through rotating cryogenic feedthroughs and minimizing
frictional contacts, this pendulum-based approach is particularly well
suited to operation with both classical gases and cryogenic liquids.

One of the key features of the CHIMERA apparatus is its ability to
operate with a wide variety of working fluids, ranging from
room-temperature gases to normal and superfluid helium. In its normal
phase, liquid helium combines an exceptionally low kinematic viscosity
($\nu \approx \SI{2.6e-8}{\meter^2 \per \second}$ at \SI{3}{\kelvin} and
$P=\SI{1}{bar}$) with favorable thermal properties, enabling extremely
large Reynolds numbers to be achieved while maintaining moderate
rotation rates, mechanical stresses and dissipated power. Such
conditions have previously allowed helium experiments to reach
Reynolds numbers among the highest ever obtained in laboratory
turbulence studies, as illustrated by the SHREK
(Superfluid High REynolds von Kármán)
experiment\cite{Rousset14_shrek}, in which turbulence is measured in a
gas or subcooled liquid helium between two counter-rotating or co-rotating
turbines.

Below the lambda transition ($T<T_\lambda$), the same apparatus can be operated in the superfluid phase (He~II), allowing direct comparisons between classical and quantum turbulent transport without changing either the geometry or the measurement principle. The combination of identical boundary conditions, identical forcing and the same global torque diagnostic provides an experimental framework particularly well suited to investigating the influence of the superfluid component on wall-bounded turbulent momentum transport.

In this article, we present the mechanical design of the pendulum-suspended outer cylinder, the optical torsion measurement system and the calibration procedures used to achieve high torque resolution over a wide temperature range. Measurements performed in gaseous helium and normal liquid helium validate the instrument and demonstrate its capability to measure turbulent torque over more than five decades in Reynolds number. The design also provides a straightforward route towards future measurements at even higher Reynolds numbers and, ultimately, to systematic comparisons between classical and superfluid Taylor-Couette turbulence using a single, precisely characterized experimental platform.

\section{\label{sec:expe}Experimental Setup}

\begin{figure*}
  \includegraphics[]{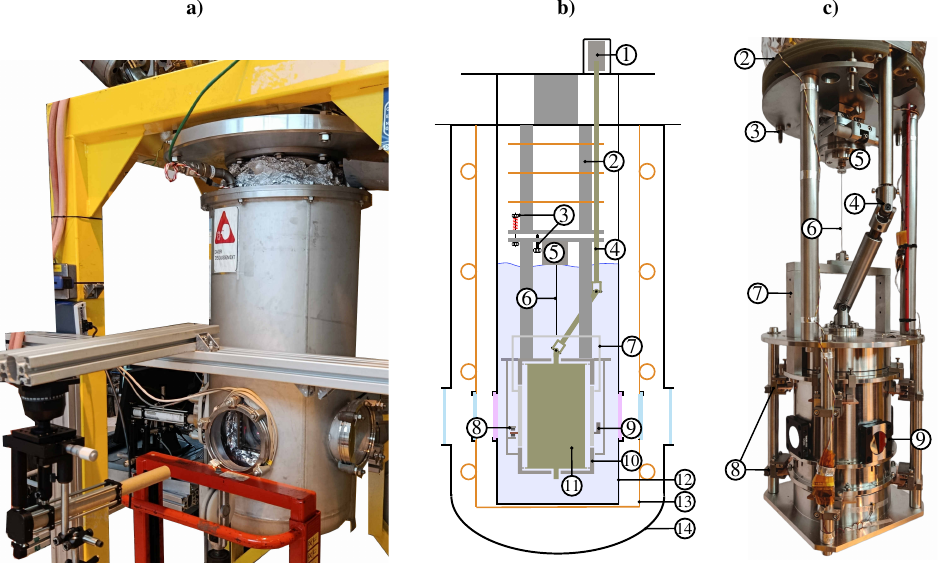}
  \caption{\label{fig:overall} a) Picture of the cryostat and
    the laser diode/PSD setup. b) Detailed schematic of the 
    Taylor-Couette apparatus. 1-Brushless DC motor, 2-Base structure of the
    cryostat, 3-Vertical alignment ($\varphi$) system, 4-Double universal joint
    shaft, 5-($x, y, z, \theta$) alignment system, 6-Stainless steel
    torsion wire, 7-Moving part of the outer cylinder with its supporting arm,
    8-Eddy-current brake, 9-Angle sensing mirror, 10-Fixed part of the
    outer cylinder, 11-Inner cylinder, 12-Helium vessel, 13-Radiation
    screen, 14-Vacuum vessel. c) Photograph of the inner part of the apparatus.}
\end{figure*}

The present system has been designed to fit within the OGRES cryostat, 
whose large optical access and temperature-control capabilities make it
particularly well suited to the present Taylor-Couette experiments (see Fig.~\ref{fig:overall}-a).

\subsection{Description of the optical cryostat}

The OGRES cryostat has been described in detail by Bret et al.~\cite{Bret25}. Here, we summarize the characteristics relevant to the present experiment, in particular the optical access, temperature control, and operating range in liquid helium.

The cryostat consists of two concentric stainless steel vessels separated by a vacuum-insulated space. To reduce radiative heat transfer, a copper heat shield equipped with multilayer insulation (MLI) is placed between the two vessels and cooled to approximately \SI{100}{K} using liquid nitrogen. Since the main objective of the experiment is to provide direct optical access to the interior of the cryostat, each of the three concentric vessels is equipped with four flat viewing ports, providing a square field of view of approximately \SI{10}{cm} $\times$ \SI{10}{cm}, with the ports spaced by $90^\circ$ around the circumference.

The inner vessel has a diameter of \SI{210}{mm} and can contain up to approximately \SI{600}{mm} of liquid helium when filled to two-thirds of its capacity. The temperature of the helium bath can be controlled by adjusting the pressure inside the vessel, allowing measurements to be performed in both He~I and He~II. Below the $\lambda$ transition, the pressure regulation provided by the valve and pressure sensor is better than \SI{1}{Pa}. This corresponds to a temperature variation of approximately \SI{0.08}{mK} at \SI{2.2}{K} and \SI{0.4}{mK} at \SI{1.5}{K}.

Our pumping system allows the temperature to be continuously reduced down to \SI{1.5}{K}. The minimum temperature is determined by the heat load and the performance of the pumping system. Table~\ref{tab} summarizes the saturation pressure, saturation temperature, superfluid fraction $\rho_s/\rho$, and sensitivity of the saturation temperature to pressure variations   for representative He~II conditions accessible in our facility. For example, at \SI{1.6}{K}, approximately 83\% of the helium density is in the superfluid component, leaving about 17\% in the normal component.

\begin{table}
\caption{\label{tab}
Saturation pressure, saturation temperature, superfluid density fraction , sensitivity of the saturation temperature to pressure variations  of He~II  for representative operating conditions accessible in the present facility.}
\centering
\begin{ruledtabular}
\begin{tabular}{c|c|c|c}
\noalign{\vskip 2pt}
$P_{\mathrm{sat}}$ [\si{\pascal}] & $T_{\mathrm{sat}}$ [\si{\kelvin}]
  & $\rho_s/\rho$ [\si{\percent}] & $ \left. \frac{\partial
                                    T}{\partial P}
                                    \right|_{\mathrm{sat}}$
                                    [\si{\milli\kelvin \per \pascal}]\\
\noalign{\vskip 2pt}
\hline\hline
472 & 1.50 & 88.3 & 0.44\\
746 & 1.60 & 83.0  & 0.31 \\
1128 & 1.70 & 76.2 & 0.23\\
1638 & 1.80 & 67.8 & 0.17\\
2299 & 1.90 & 57.2 & 0.14\\
3129 & 2.00 & 43.6 &  0.11\\
4141 & 2.10 & 24.8 & 0.091  \\
4597 & 2.14 & 14.4 &  0.085\\
\end{tabular}
\end{ruledtabular}
\end{table}

Cernox\textsuperscript{\textregistered} thermometers are positioned at
different levels inside the cryostat.  
Their accuracy improves as the temperature decreases and remains
better than \SI{5}{mK} even at \SI{3}{K}. In He~II,  
the thermometers give identical readings, as the fluid is essentially
isothermal.  
For experiments in He~I, it is necessary to operate under subcooled conditions (see Section~\ref{sec:MeasurementProcedure}). 
The two thermometers located at the upper and lower ends of the
Taylor-Couette flow cell are then used to monitor the spatial and 
temporal evolution of the temperature and to ensure that the
temperature remains sufficiently homogeneous throughout the
experiment: typically they indicate the same value at the start of an
experiment, just after represurization, then the observed drift is
faster for the upper one ($\approx\SI{3}{\milli\kelvin\per\minute}$)
than for the lower one ($\approx\SI{2}{\milli\kelvin\per\minute}$).

\subsection{Overall Description}
The Taylor-Couette flow cell is made of two concentric
cylinders of which  only the inner cylinder can be set in
rotation, the outer being at rest with respect to the rest of the
cryostat. The design of the apparatus was specifically optimized to study
the mean momentum flux between the two cylinders through the measurements
of the induced torque, in the widest possible range of flow regimes.

\subsubsection{Geometry}
The entire system is built with the same grade of stainless steel 
(X2CrNi18-9) to ensure that all proportions are conserved after
cool-down to cryogenic temperature. The lengths given below are
measured at room temperature and the expected contraction at
the lowest temperature is of the order of
\SI{0.3}{\percent}\cite{VanSciver1986}. 

The chosen geometry is as follows: the radii of the inner
and outer cylinders are $R_i = \SI{50\pm0.075}{\milli\meter}$ and
$R_o= \SI{59\pm0.075}{\milli\meter}$  
respectively (with their cylindricity tolerance better than
\SI{0.09}{\percent}), and the total height $H = \SI{196\pm
  0.2}{\milli\meter}$. From those numbers 
we can compute classical control parameters, namely the gap
$\delta = (R_o-R_i) = \SI{9}{\milli\meter}$, the radius ratio $\eta = R_i/R_o
\approx\num{ 0.85}$, and the aspect ratio $\Gamma = H/\delta\approx
\num{22}$. The surface roughness of the cylinders walls in contact with the
flow is $Ra < \SI{0.2}{\micro\meter}$. 

Because the aspect ratio is finite, the end boundaries break the axial
homogeneity of the flow and generate boundary layers and secondary
circulations. Such end-wall effects can significantly modify the flow
structure and the local angular-momentum transport in finite-length
Taylor-Couette systems, as demonstrated experimentally by Coles and
Van Atta \cite{Coles_Atta_1966}. To limit the contribution of the end
regions to the torque measurement, we adopt an axially segmented
configuration, in which the cylinders are divided into three
sections—top, middle, and bottom—and torque is measured only on the
middle section. This approach has been used in previous Taylor-Couette
torque measurements
\cite{Lathrop92_taylor_couette_long,VanGils11_taylor_couette_twente}. Importantly,
Lim and Tan \cite{LimTan2004} showed experimentally that, for aspect
ratios $20< \Gamma < 40$, the torque per unit height becomes
essentially independent of the aspect ratio at Reynolds numbers
moderately above the onset of Taylor vortices, suggesting that
finite-aspect-ratio effects on the measured torque can become
relatively weak in this regime.  
Our apparatus has an aspect ratio $\Gamma =22$, which lies within this
range. Measuring the torque only on the central section provides an
additional precaution against any direct contribution from the end
regions. 

Contrary to most experiments, the torque is measured here on the outer 
cylinder. Its top and bottom parts (item 10 in Fig.~\ref{fig:overall}-b),
each $\SI{27\pm0.1}{mm}=3\delta$ long,
are rigidly attached to the apparatus while the middle section
 (item 7 in Fig.~\ref{fig:overall}-b),
$h=\SI{140\pm0.1}{mm}$ long, is connected to a separate alignment system. The gaps between 
the middle section and the top and bottom sections are \SI{1}{mm} at each end, giving a total clearance of \SI{2}{mm}. 
The three sections are carefully aligned so that this clearance is
distributed as evenly as possible between the two ends. 

\subsubsection{Motor}
Motors operating at low temperature generally have a low efficiency
and thus dissipate a significant
amount of heat, thereby increasing the lower limit of working temperature. 
Consequently, the inner cylinder is set in rotation by a remote brushless
DC motor (item 1 in Fig.~\ref{fig:overall}-b) situated at room temperature
in a dedicated enclosure. The torque measurement system (see Sec.~\ref{sec:torquemeas}), 
and specifically the torsion wire, is situated right above the cylinders,
on the rotation axis, so that the motor cannot
be directly aligned. Motion is therefore transmitted through a double universal-joint shaft (item 4 in Fig.~\ref{fig:overall}-b),
which allows shifting the motor
axis away from the cylinder axis. Additionally, the middle section of the shaft
is telescopic to accommodate possible differential contraction.

The cylinder rotation frequency $f$ can be controlled to better than \SI{1}{\percent}
accuracy in the range $\SI{0.3}{\hertz}<f<\SI{7.8}{\hertz}$, without adding any significant additional heat loss.

\subsection{\label{sec:torquemeas}Torque measurement system}

In what follows, we present the fundamental operating
principle of our torsion-based measurement system, which is designed
to determine the torque acting on the outer cylinder. We subsequently
provide a detailed description of the cylinder suspension, the
oscillation damping strategy, and the angular displacement measurement
techniques. 

\subsubsection{Base principle}
The most straightforward approach consists in inferring the torque
directly from the motor power consumption or from a torque meter
mounted on the rotation axis. However, this method does not permit the
subtraction of parasitic torques arising from ball bearings and end
effects of the cylinders (see, e.g.,
\textcite{Ravelet10_taylor_couette}). A significant improvement was
introduced by \textcite{Lathrop92_taylor_couette_long}, who segmented
the inner cylinder into three axial sections so as to isolate and
measure only the torque associated with the central region. In their
configuration, the torque was measured by a strain-gauge-based torque
sensor installed between the rotation axis and the middle section, and
the electrical signal was transmitted via a rotary feedthrough. A
similar strategy has been adopted by the Twente group for their
large-scale Taylor-Couette
device~\cite{VanGils11_taylor_couette_twente}. This technique was
further refined by \textcite{Butcher24_taylor_couette_okinawa}, who
eliminated the noise introduced by the rotary feedthrough by
digitizing the analogue output of the torque meter prior to its
transmission through the slip ring.

Considering the difficulty of implementing such instrumentation in a cryogenic
environment, we developed a completely different measurement approach which,
among other advantages, avoids the need for electrical feedthroughs and
low-temperature signal acquisition electronics. Furthermore, it also
eliminates the static torque associated with the ball bearing between
the rotation shaft and the sensing section of the inner cylinder, allowing
the lower limit of measurable torques to be further reduced.

Our design was inspired by early low-temperature Couette viscometers used to
measure the viscosity of superfluid helium (He~II), in which the inner cylinder
was suspended from a torsion wire while the outer cylinder was driven in
rotation~\cite{Hollis53_taylor_couette_helium}. In our case, however, the inner
cylinder must be actively driven, which makes such a suspension impractical.
For this reason, we reversed the principle of these historical experiments and
instead measure the torque exerted by the flow on the outer cylinder.

\subsubsection{Outer cylinder suspension}
The outer cylinder is suspended from the rest of the cryostat by a cylindrical
stainless steel wire. Its main axis naturally aligns with the vertical direction
under gravity. It is therefore necessary to match the vertical axis of
the rest of the apparatus with that of the outer cylinder. 

For this purpose, the bottom part of the apparatus is attached to the main
cryostat structure through a system of three springs compressed by adjustable
stops (item~3 in Fig.~\ref{fig:overall}-b).

The central part of the outer cylinder (item 7) is horizontally with
the top and bottom sections  using micrometer-controlled adjustment
stages  (Misumi E-XYSSG40) 
located at the top of the suspension wire (item 5), allowing precise adjustment 
of the relative position of the suspended section in the transverse directions.

This vertical and horizontal alignment procedure is repeated until we 
estimate that the gaps above and below the middle section are identical at 
every point on the circumference of the cylinders, their respective 
heights being equivalent, and the horizontal alignment is
also better than \SI{0.1}{mm} accuracy.

During cryogenic operation with liquid helium, the alignment is
monitored visually through the cryostat viewing ports. In particular,
the position of the copper plates within the eddy-current brake
assemblies provides a sensitive indication of any significant
displacement of the suspended cylinder.

\subsubsection{Angle measurement}

\begin{figure}
  \includegraphics[]{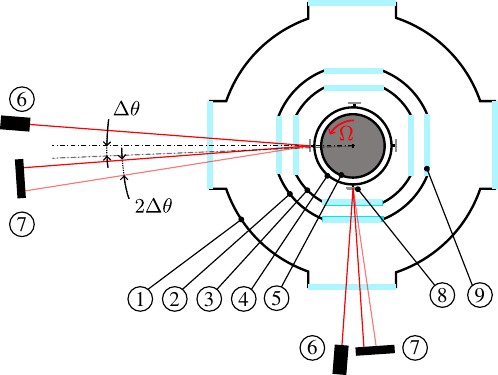}
  \caption{\label{fig:psd} Top view schematic of the torsion angle
    measurement system. 1-~Vacuum vessel, 2-~Radiation screen, 3-~Helium vessel,
    4-~Outer cylinder, 5-~Inner cylinder, 6-~Laser diodes, 
    7-~2-D position sensing devices, 8-~Mirrors, 9-~Windows}
\end{figure}

Figure ~\ref{fig:psd} presents a detailed schematic of the torque
measurement system.

A light beam emitted by a \SI{1}{\milli\watt} laser diode 
(Laser Components\textsuperscript{\tiny\textregistered} 
 FP-D-635-1-C-F, item~6 in
Fig.~\ref{fig:psd}) passes through the three windows of the cryostat
before being 
reflected by a mirror attached to the outer cylinder. The reflected beam then
travels back through the windows and impinges on the surface of a
two-dimensional $\num{10}\times\SI{10}{\milli\meter^2}$
position-sensitive device (PSD, SiTek SEEPOS\textsuperscript{\tiny\textregistered}
2L10\_MH01, item~7).
When the inner cylinder (item~5) is set in rotation, the fluid transports
angular momentum to the outer cylinder (item~4), which is then subjected
to a torque $T$ and rotates, together with the attached mirrors, by a
small angle $\Delta \theta = T/k$ about its axis, where $k$ is the torsion
(spring) constant of the suspension wire. The laser beam is consequently deflected
by an angle $2\Delta \theta$, and the resulting displacement $\Delta x$
measured on the PSD is, at first order, given by
$$
\Delta x = 2 L_{\mathrm{PSD}} \Delta \theta,
$$
where $L_{\mathrm{PSD}}$ is the distance between the mirror and the surface
of the PSD.

The advantage of using a two-dimensional PSD is that it allows us not only to measure rotations around the vertical axis, which produce horizontal displacements $\Delta x$, but also to detect possible vertical displacements caused by spurious swinging of the outer cylinder.

Two such systems, with different distances $L_{PSD-1}\approx\SI{773}{\mm}$ and 
$L_{PSD-2}\approx\SI{435}{\mm}$, are used simultaneously, allowing for an increased
range of measurable torques without losing accuracy.

\subsubsection{Oscillation damping system}

\begin{center}
  \begin{figure}
    \includegraphics[]{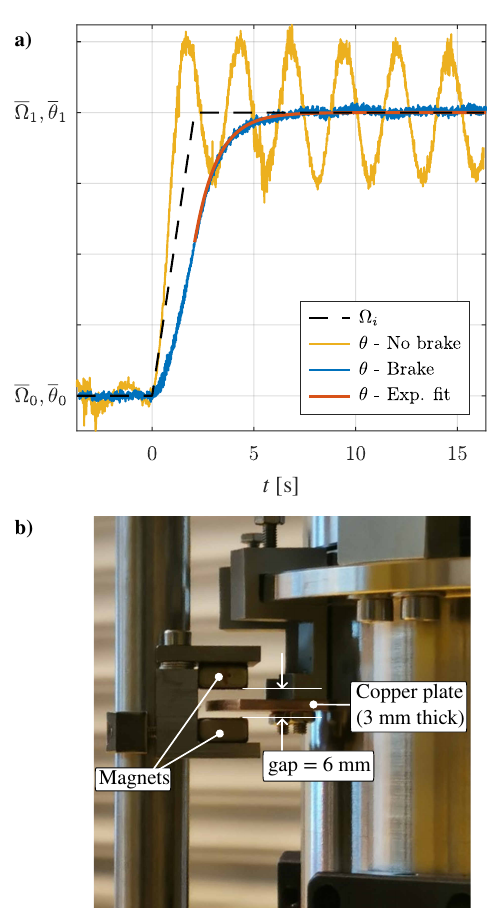}
    \caption{\label{fig:eddybreak} 
      a) Torsion angle $\theta$ evolving from an
      equilibrium position to another (denoted $\overline{\theta}_0$
      and $\overline{\theta}_1$
    here) after a change in the rotation frequencies $\Omega_i$ of the
    inner cylinder (black dashed line) from  $\overline{\Omega}_0$ to
    $\overline{\Omega}_1$. The blue and yellow solid 
    lines correspond to measurements of $\theta$ performed at room temperature 
    with and without the eddy-current brake respectively. The orange
    solid line is an exponential fit of the blue curve
    after the acceleration period.
      b) Annotated picture of the eddy-current brake setup.}
  \end{figure}
\end{center}

Out of equilibrium, such systems produce large oscillations about the
torsion axis that are only weakly damped over time (see yellow curve in
Fig.~\ref{fig:eddybreak}-a): the typical oscillation period at room
temperature is of the order of  
\SI{3}{\second}, while the damping time is of the order of thousands
of seconds in air.

Eight eddy-current brakes, four at the top of the outer cylinder and
four at the bottom,  have been added to fully damp these
oscillations. Each brake consists of two pairs of  $\mathrm{NdFeB}$
magnets ($\SI{10}{mm}\times\SI{10}{mm}\times\SI{5}{mm}$ cuboids), 
attached to the fixed structure and facing each other, separated by a gap of
\SI{6}{mm}, thereby producing a vertical static magnetic field.
A \SI{3}{mm} thick copper plate, attached to the outer cylinder (see
Fig.~\ref{fig:eddybreak}-b), can move horizontally in the gap,
perpendicularly to the magnetic field.

The motion of the plate in one direction, perpendicular to the magnetic field,
produces a torque $T_b$ in the opposite direction, 
proportional to the angular velocity, $T_b =- b\dot{\theta}$, where
the brake efficiency 
scales as $b \propto B^2/\rho_e$, with $B$ the magnetic field and $\rho_e$ the electrical
resistivity of the copper plate. The temporal evolution of the torsion angle $\theta$
can therefore be written as
\begin{equation}
    \ddot{\theta} + 2\omega_b \dot{\theta} + \omega_0^2 \theta = T/I ,
\end{equation}
where $2\omega_b = b/I$, $I$ being the moment of inertia of the outer cylinder about its
rotation axis, $\omega_0 = \sqrt{k/I}$ the natural angular frequency of the oscillator,
and $T$ the forcing torque.

As long as $\omega_0 < \omega_b$, after a change in the forcing torque, the system
relaxes exponentially toward a new equilibrium torsion angle, effectively suppressing
any oscillation. The relaxation time $\tau_r$ is given by
\begin{equation}
    \tau_r = \left[\omega_b - \sqrt{\omega_b^2 - \omega_0^2}\right]^{-1}.
\end{equation}

At room temperature, the typical relaxation time is of the order of
$\tau_r \approx \SI{1.1}{s}$ (see exponential fit in
Fig.~\ref{fig:eddybreak}-a), 
while it is of the order of $\tau_r \approx \SI{56}{s}$ at low temperature.
From these values, one can determine
$\omega_b = (\tau_r \omega_0^2 + 1/\tau_r)/2$ and, assuming that the natural
angular frequency of the system does not vary significantly, obtain the ratio
of the brake efficiency at these temperatures,
$r = b(\SI{4}{K})/b(\SI{300}{K}) \approx 45$.

Assuming that the magnetic field is not significantly affected
(\textcite{Diez18_cold_magnets} measured changes of the order of \SI{5}{\percent}),
this ratio of brake efficiencies provides an estimate of the electrical resistivity ratio
$r \sim \rho_e(\SI{4}{K})/\rho_e(\SI{300}{K})$. The latter depends strongly on the
grade of copper used and on impurity levels, and a value of about 45 is in the range
of observed values -- typically from 5 to 2000 (see, e.g., \textcite{VanSciver1986}).

\section{Measurements}
In this section we first describe the calibration and measurement
procedure and then  the first results obtained in our apparatus.

\subsection{\label{sec:calib}Calibration}
In order to properly infer the torque from the deviation of the laser
beam $\Delta x$, the spring constant $k$ has to be determined
precisely.

In general, such calibration is performed using known weights attached at a known
radius of the cylinder~\cite{Lathrop92_taylor_couette_long, VanGils11_taylor_couette_twente, Butcher24_taylor_couette_okinawa}, hanging under a pulley. However, the latter 
(even one with low friction) introduces a static torque, preventing low-torque
measurements. This is particularly problematic in our case, as the smallest torques measured in our experiments are as low as $\SI{1}{\micro N\cdot m}$.

Here, we use a different calibration method based on measurements of the natural frequency of the system, which overcomes this limitation.

\begin{figure}
  \includegraphics[]{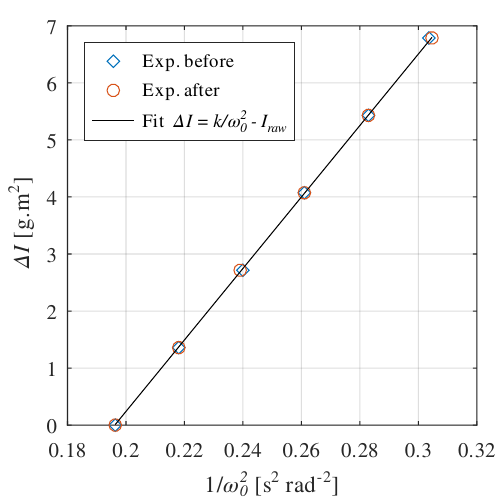}
  \caption{\label{fig:inertia} Additional inertia $\Delta I$ as a
    function of $1/\omega_0^2$. Open symbols represent calibration
    data measured at room temperature, before (blue diamonds) and
    after (red circles) cool-down to cryogenic
    temperatures, and the solid line is a linear fit with $k\approx
    \SI{63.1e-3}{\newton.\meter/\radian}$ and
    $I_{raw}\approx\SI{12.4}{g.m^2}$.}
\end{figure}

In the absence of eddy-current brakes, azimuthal oscillations of the
outer cylinder are weakly damped, the approximate relaxation time being of the order of \SI{2800}{\second},
and the natural angular frequency $\omega_0 = \sqrt{k/I}$ can be
measured very accurately. Since the moment of inertia is difficult to
compute accurately, we used the following procedure to determine it
experimentally together with the spring constant. We gradually added
calibrated masses at a known distance $R$ from the outer cylinder torsion
axis on both sides and measured the evolution of the natural angular
frequency of the system. The resulting moment of inertia is then 
$$I = I_{raw} +\underbrace{2nmR^2}_{\Delta I},$$
where $n$ is the number of added mass pairs and $I_{raw}$ is the
moment of inertia of the bare system (plus arms to support mass pairs). 
The added moment of inertia $\Delta I$ can then be rewritten
\begin{equation}
  \label{eq:fit}
  \Delta I = k/\omega_0^2- I_{raw}.
\end{equation}

In Fig.~\ref{fig:inertia} we have plotted $\Delta I$ as a function of
$1/\omega_0^2$, measured at room temperature before the first cool-down 
and after warm-up, from which a fit gives
$I_{raw}\approx\SI{12.4(0.04)}{g.m^2}$
(and \SI{10.6(0.04)}{g.m^2} for the bare system) 
 and $k\approx \SI{63.1(0.2)}{\milli\newton.\meter\per\radian}$.

From this value, one can compute the shear modulus $G$ of the
stainless steel torsion wire of length $L_w=\SI{115\pm 1}{\milli \meter}$
and diameter $D_w\approx \SI{0.969\pm 0.003}{\milli \meter}$:
$$G = \frac{32kL_w}{\pi D_w^4}\approx \SI{84\pm 1.3}{\giga\pascal}.$$

For comparison, the shear modulus of bulk stainless steel in its various
grades is expected to lie in the range
$G \approx \SI{75\pm 3}{\giga\pascal}$ \cite{Ledbetter81_youngmodulus}.
The measured value is \SI{11}{\percent} larger which may be explained
by the hardening of the wire material (actually a 316L soldering rod)
during its manufacturing.

A dedicated run has been performed at cryogenic temperature to estimate
the spring constant for liquid helium experiments, $k_{\SI{2}{\kelvin}}$,
from the ratio of the natural frequency of the oscillations
at room temperature and low temperature:
$$k_{\SI{2}{\kelvin}} = k
\left(\frac{\omega_{0-\SI{2}{\kelvin}}}{\omega_{0}}\right)^2 \frac{I_{\SI{2}{\kelvin}}}{I}.$$
We measured
$\omega_{0-\SI{2}{\kelvin}}/\omega_{0}\approx1.027$
and, assuming that the contraction of the stainless steel is about
\SI{0.3}{\percent}\cite{VanSciver1986}, we obtained
$I_{\SI{2}{\kelvin}}/I\approx 0.994$ and then
$k_{\SI{2}{\kelvin}} = 1.048 k$.
This increase of \SI{4.8}{\percent} is smaller than reported by
\textcite{Ledbetter81_youngmodulus}, who measured an increase of
approximately 8\% for stainless steel between room temperature and low
temperatures. Again, the crystallographic state of the metal may play
an important role. 
Like \textcite{Ledbetter81_youngmodulus}, we did not observe
significant changes between \SI{2}{\kelvin} and \SI{3}{\kelvin} and we
consequently used a single value $k_{\SI{2}{\kelvin}}$ for 
measurements at all cryogenic temperatures.

\subsection{Measurement procedure\label{sec:MeasurementProcedure}}
A systematic torque measurement procedure has been established with
the aim of detecting possible drifts in the position signal delivered
by the PSD. For this purpose, upward then downward speed ramps are
programmed, including certain redundant points, and intermediate zero
measurements. Note that this procedure is dedicated to the validation
of the apparatus and its measurement principle.
In order to make sure that the state
that is reached after a change in rotation rate is the same  as the
previous one (notably the number of Taylor-vortex cells in the
intermediate Reynolds number range $10^2<Re_s<10^5$), a more refined
procedure, including controlled acceleration, should be applied in the
future (see, e.g., \textcite{coles1965transition}).

Regarding temperature and pressure control, three measurement
cases must be distinguished:  gas at room temperature,
normal liquid helium (He~I), and finally superfluid helium (He~II).

At room temperature, the cryostat is first emptied using a primary
vacuum pump down to $P\approx \SI{0.25}{\milli\bar}$, and then
pressurized again up to the target pressure $P\leq\SI{1}{\bar}$
using the chosen gas. Possible temperature and pressure drifts 
during a measurement cycle (lasting approximately one hour) are
monitored in order to account for any variations in the resulting
kinematic viscosity calculation (see
Sec.~\ref{sec:errors}). Our measurements show that the heat capacity of the
apparatus is sufficient to ensure that the temperature variation
during a measurement cycle is negligibly small.

At cryogenic temperature, and in the presence of heat inputs, ensuring
stable pressure and temperature  during a measurement cycle is much
easier in He~II than in He~I.  
Indeed, in He~II, working at saturation pressure and temperature
is not an issue since very  efficient heat transfer mechanisms
homogenize the bath temperature and prevent the appearance of
bubbles. In that case, a pressure control 
system allows us to ensure  pressure stability better than \SI{1}{\pascal} (and temperature stability better than \SI{0.4}{\milli\kelvin} down to \SI{1.5}{K}). 
On the other hand, in He~I, working at saturation is not an option
because parasitic bubbles would appear. The procedure is to first
pump the liquid helium bath down to the target pressure (temperature)
and then instantly repressurize up to \SI{1}{\bar} (the maximum
allowable pressure in this cryostat). This results in a 
stratification of the liquid: the temperature at the free surface
rapidly reaches the saturation ($\approx \SI{4.2}{\kelvin}$),
while the temperature of the rest of the liquid remains mostly unchanged,
as the liquid there is denser. At that point measurements can
be started, and the slow temperature drift due to heat inputs is recorded
and subsequently used to  determine the actual kinematic viscosity of the liquid.
\subsection{First results}

  \begin{figure*}
    
  \includegraphics[]{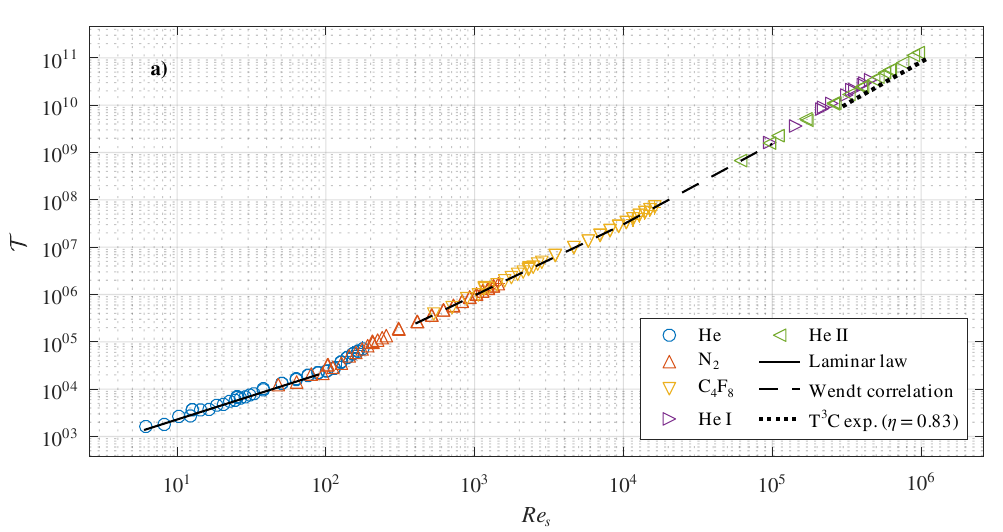}\\
  \includegraphics[]{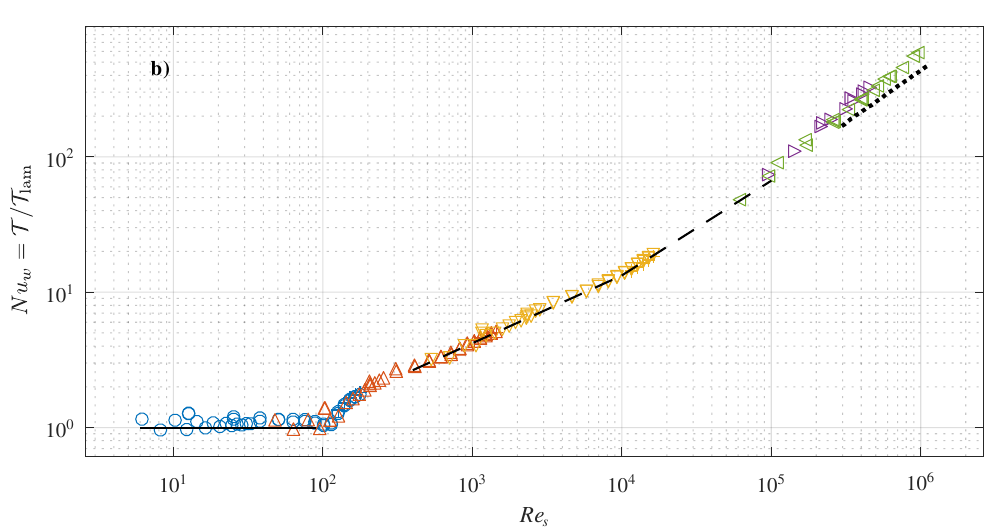}
  \caption{\label{fig:result} a) Nondimensional torque as a function
    of the Reynolds number for various gases
    ($\mathrm{He}$, $\mathrm{N_2}$ and $\mathrm{C_4F_8}$) at
    room temperature and pressures of \SI{0.3}{\bar} and \SI{1}{\bar},
    for liquid helium He~I at 2.2 and \SI{3.4}{\kelvin}, and for liquid helium
    He~II at 1.6 and \SI{2.14}{\kelvin}. The kinematic  viscosity of
    He~II is taken to be the ratio between the normal component's 
    dynamic viscosity and the total density. For comparison, we added the
    laminar theoretical prediction $\mathcal{T}_{\text{lam}}$ 
    (solid line, $6\leq Re_s \leq 90$), the Wendt
    correlations\cite{wendt1933turbulente} (dashed line, $400\leq Re_s
    \leq 10^5$) and the data from the 
    T$^3$C experiment\cite{Grossman16_tc_high_re} (dotted line, $\num{3e5}\leq
    Re_s \leq \num{1.1e6}$) at a slightly different $\eta$ (0.83 vs
    0.85).
    b) Same data divided by the nondimensionalized theoretical laminar
    torque -- the so called Nusselt number for Taylor-Couette flows.}
\end{figure*}
We have used the apparatus to measure the torque in a very wide range
of Reynolds numbers.
Because we want to make use of liquid helium, our experiment has to be enclosed
in a sealed vessel, allowing us to also use several gases with
different kinematic viscosities, as is customary in some wind
tunnels\cite{Bodenschatz14_uboat,Zagarola97_princeton_pipe}. In our
case, as the vessel cannot be pressurized above \SI{1}{\bar},
$\mathrm{C_4F_8}$ is the best candidate at ambient pressure and
temperature for obtaining the lowest  possible kinematic viscosity in
a gas, while gas helium is the fluid with the  highest kinematic
viscosity. Finally, the kinematic viscosity at ambient 
temperature can be further increased by lowering the pressure of the
gas used. To summarize, we used gas $\mathrm{He}$, $\mathrm{N_2}$ and
$\mathrm{C_4F_8}$, at room temperatures (at \SI{300}{K}, $P =
\SI{0.3}{bar}$ or $\SI{1}{bar}$), which allowed us to cover  
more than three decades of Reynolds numbers, from \num{6} to
\num{1.6e4}, and then liquid helium  at temperatures between
\SI{1.6}{K} and \SI{3.6}{K}, which covers the highest 
values $\num{6e4}< Re_s < \num{1e6}$.
For He~II experiments  the fluid is described using the two-fluid
model\cite{Tisza_two_fluids_38, Landau_two_fluids_41}, and
the choice of an effective viscosity is not straightforward. In
the present work, we define the kinematic viscosity as the ratio
of the dynamic viscosity of the normal component to the total
density of the fluid. With this definition, the experimental data
exhibit a consistent scaling behavior over the explored range
of parameters. While this empirical choice appears to provide
a relevant characterization of the global transport, its physical
justification remains an open question and will be discussed
in more detail in a dedicated study.

Figure~\ref{fig:result}-a shows the dimensionless torque $\mathcal{T}
= T/h\rho\nu^2$ as a function of the Reynolds number.
In the spirit of the Rayleigh-Bénard convection analogy,
Fig.~\ref{fig:result}-b further shows the equivalent of the Nusselt
number for Taylor-Couette flows, i.e. the nondimensional torque divided by its 
expected value  in purely diffusive (laminar) regime : $$Nu_w =
\mathcal{T}/\mathcal{T}_{\text{lam}},$$
where $\mathcal{T}_{\text{lam}}= 2\pi Re_s \eta/(1-\eta)^2$.


Below the first critical Reynolds number $Re_c\approx 100$ the data,
obtained with gas  helium and gas Nitrogen  follow very well the
analytical (solid black line) law, confirmed by the approximately
constant value of $Nu_w = \num{1.1\pm 0.1}$.

After a transitional region the data for $Re\gtrsim 400$, from Nitrogen and then
$\mathrm{C_4F_8}$ compare very well with 
the correlation proposed by Wendt\cite{wendt1933turbulente} (dashed black line),
i.e., a power law
$\mathcal{T}\propto Re^{1.5}$ for $Re< 10^4$ and then a transition to a steeper power law
$\mathcal{T}\propto Re^{1.7}$. At very large Reynolds number, our data are compared to
those obtained by the group of Twente (extracted from
\textcite{Grossman16_tc_high_re}) 
at $\eta=0.83$, the closest to ours ($\eta=0.85$). 
Again, the torque behavior in these highly turbulent Taylor–Couette regimes compares reasonably well with the T$^3$C data.



\subsection{\label{sec:errors}Uncertainty on the torque}
\begin{table}[h]
\caption{\label{tab:errors}Main contributions to the torque
  uncertainty. The minimum and maximum relative uncertainties $\Delta
  T/T_0$ are  computed for the largest and smallest measured values of
  the torque $T_0$ respectively.}
\begin{ruledtabular}
\begin{tabular}{l||c|c|c}
  Source
  & $\Delta x  [\si{\micro\meter}]$
  & $\Delta T [\si{\micro\newton\meter}]$
  & $\Delta T/T_0$ [\si{\percent}]\\
  & & & min -- max\\
  \hline\hline

  Vibrations 
  & $\num{4.5}$ 
  & $\num{0.2}$
  & 0.8 -- 15\\

  Speckle (He I) 
  & 20 
  & 0.9 
  & 0.3 -- 4\\

  Zero drift (He I) 
  & \num{50} 
  & \num{2.3}
  & 0.8 -- 10\\

\end{tabular}
\end{ruledtabular}
\end{table}

Several sources of uncertainty affect the torque measurements, with
their relative importance depending on the working fluid and operating
conditions (see Tab.~\ref{tab:errors} for a summary). The first
contribution is associated with residual pendular 
oscillations of the suspended outer cylinder. These oscillations correspond
to a swinging motion about the vertical axis and can be detected through
the vertical displacement of the laser spot on the PSD. No clear
correlation is observed between the vertical and horizontal signals,
indicating that the contribution to the torque signal cannot be described
as a simple geometrical projection of the pendular motion. Nevertheless,
the resulting fluctuations contribute to the short-time noise of the
torque measurement. Their amplitude is of the order of
$\Delta x_{\mathrm{vib}}\approx\SI{4.5}{\micro\meter}$ at the highest
rotation frequencies investigated. This corresponds to apparent torque
fluctuations of up to
$\Delta T_{\mathrm{vib}}\approx\SI{0.2}{\micro\newton\meter}$. 
Note that since this corresponds to a short time pseudo-periodic noise,
averaging over a duration much larger than the noise period
effectively minimizes the resulting uncertainty. The
associated relative uncertainty reported in Tab.~\ref{tab:errors}
is thus a very pessimistic value computed assuming that no averaging
is performed.

A second contribution is associated with slow fluctuations of the
zero-torque reference. These low-frequency drifts are observed in
liquid-helium measurements and can occur on time scales of several tens
of minutes. Their amplitude can reach approximately
$\Delta x_{\mathrm{drift}}\approx\SI{50}{\micro\meter}$, corresponding
to $\Delta T_{\mathrm{drift}}\approx\SI{2.3}{\micro\newton\meter}$.
The zero-torque position is therefore recalibrated regularly during the
measurements in order to minimize their effect. This procedure reduces
the resulting error, although slow variations of the zero reference
necessarily introduce a residual bias in the measured torque.

Finally, optical speckle provides an additional source of uncertainty
in liquid helium. This effect is particularly pronounced in He~I, where
refractive-index fluctuations and multiple reflections through the
cryostat and the fluid produce a dispersion of the beam barycenter on
the PSD. The resulting positional uncertainty is of the order of
$\Delta x_{\mathrm{speckle}}\approx\SI{20}{\micro\meter}$, corresponding
to $\Delta T_{\mathrm{speckle}}\approx\SI{0.9}{\micro\newton\meter}$.
The speckle contribution is strongly reduced in He~II, where the optical
signal is substantially cleaner.

The relative importance of these contributions therefore depends strongly
on the operating conditions. In gaseous helium, the pendular contribution
is the dominant short-time noise source,
whereas the zero reference remains stable over the duration
of a measurement. In liquid helium, the optical speckle and slow
zero-reference drifts become the dominant additional contributions,
particularly in He~I.

\section{Conclusion}


We have presented  a Taylor-Couette apparatus designed to investigate angular momentum transport over an extended range of Reynolds numbers, from classical gaseous flows to cryogenic helium, including the superfluid regime. The instrument relies on a torsion-pendulum configuration in which the outer cylinder serves as a direct torque sensor, thereby avoiding torque transmission through rotating mechanical elements and enabling operation in demanding cryogenic environments.

The combination of optical angle detection using a two-dimensional position-sensitive detector and efficient eddy-current damping provides a sensitive and robust measurement of the steady-state torque, with a well-controlled temporal response. A calibration procedure based on the measurement of the natural oscillation frequency allows for an accurate determination of the torsion constant without introducing additional parasitic torques.

First measurements performed in various gases and in liquid helium demonstrate that the apparatus is capable of spanning more than five decades in Reynolds number, up to $Re_s \sim 10^6$.  These results validate both the measurement principle and the overall mechanical design.

Beyond its performance as a high-resolution torque sensor, the ability
to operate the same setup across the normal and superfluid phases of
helium opens new perspectives for the study of turbulent transport in
quantum fluids. In particular, the present instrument provides direct
access to the global angular momentum flux, offering a promising route
to investigate how quantized vortices and mutual friction contribute
to turbulent dynamics and scaling laws.

This facility thus constitutes a versatile experimental platform for exploring Taylor-Couette turbulence across a wide range of physical regimes, bridging classical and quantum fluid dynamics within a unified framework.

\begin{acknowledgments}
We acknowledge the financial support of the Cross-Disciplinary 
Program on Instrumentation and Detection of CEA, the French Alternative Energies
and Atomic Energy Commission.

We are very grateful  to  Florian Bancel for his help in the design of the apparatus,
as well as Bérengère Dubrulle, Adam Cheminet and Jules Delacroix for
fruitful discussions.
\end{acknowledgments}

\section*{Data Availability Statement}

AIP Publishing believes that all datasets underlying the conclusions of the paper should be available to readers. Authors are encouraged to deposit their datasets in publicly available repositories or present them in the main manuscript. All research articles must include a data availability statement stating where the data can be found. In this section, authors should add the respective statement from the chart below based on the availability of data in their paper.

\bibliography{RSI_streams.bib}

\end{document}